\documentclass[AMA,Times2COL]{WileyNJDv5}

\makeatletter

\def\oddhead@titlepage@info{}
\def\evenhead@titlepage@info{}
\def\oddfoot@titlepage@info{}
\def\evenfoot@titlepage@info{}

\makeatother

\articletype{Empirical}%

\journal{}

\begin{document}

\title{A Taxonomy of Runtime Faults in Model Context Protocol Servers}

\author[1,2]{Joshua Owotogbe}
\author[1,2]{Indika Kumara}
\author[4]{Antonio Ken Iannillo}
\author[1,2]{Willem-Jan van den Heuvel}
\author[1,3]{Damian Andrew Tamburri}
\author[5]{Roberto Natella}

\authormark{OWOTOGBE \textsc{et al.}}

\titlemark{A Taxonomy of Runtime Faults in Model Context Protocol Servers}

\address[1]{%
\orgname{Jheronimus Academy of Data Science (JADS)},
\orgaddress{
\city{'s-Hertogenbosch},
\state{North Brabant},
\country{The Netherlands}}}

\address[2]{%
\orgname{Tilburg University},
\orgaddress{
\city{Tilburg},
\state{North Brabant},
\country{The Netherlands}}}

\address[3]{%
\orgname{University of Sannio},
\orgaddress{
\city{Benevento},
\country{Italy}}}

\address[4]{%
\orgname{University of Luxembourg},
\orgaddress{
\city{Luxembourg},
\country{Luxembourg}}}

\address[5]{%
\orgname{Gran Sasso Science Institute},
\orgaddress{
\city{L'Aquila},
\country{Italy}}}

\corres{Corresponding author: Joshua Owotogbe,
Jheronimus Academy of Data Science (JADS) and Tilburg University,
The Netherlands.
\email{j.s.owotogbe@tilburguniversity.edu}}


\abstract[Abstract]{MCP (Model Context Protocol) enables LLMs (Large Language Models) to interact with external tools and data sources via a standardized protocol. Its rapid adoption in tool-augmented Artificial Intelligence (AI) workflows has introduced new reliability challenges, such as configuration parameters that are accepted but not enforced at runtime, leading to unintended default behavior,  whose runtime fault characteristics remain empirically unexamined. We present the first empirical taxonomy of runtime faults in MCP servers. We manually analyzed 833 confirmed runtime fault threads identified through a repository-mining pipeline covering 473 actively maintained MCP server GitHub repositories and derived a taxonomy using a bottom-up open coding procedure. The taxonomy comprises 11 top-level categories and 27 subcategories (73 leaf fault types), covering recurrent failures across protocol interactions, tool invocations, schema enforcement, state management, model-provider integration, security validation, and timeouts or explicit cancellations of in-progress operations. To assess the taxonomy's external validity, we surveyed 55 MCP server developers. Respondents reported experiencing an average of 20 of the 27 fault subcategories, and no category remained unobserved. These results indicate that the taxonomy reflects widely observed runtime failures in MCP-based systems and shall assist AI software maintenance and evolution in the future.}



\maketitle

\renewcommand{\thefootnote}{\arabic{footnote}}
\setcounter{footnote}{0}

\section{Introduction}
\label{sec:introduction}

Large language models (LLMs) have enabled automation of complex tasks across domains such as software development, legal, and customer support~\cite{zhao2023survey, bommasani2021opportunities}, yet they typically lack the domain-specific context required for real-world deployments~\cite{yang2025comprehensive, tonmoy2024comprehensive} and must be augmented with data from external sources such as APIs and databases~\cite{lewis2020retrieval, gao2023retrieval, yao2022react, schick2023toolformer}. Connecting AI models to existing organizational systems, however, requires complex integration and specialized skills, creating a scalability bottleneck as the number of models and external systems grows~\cite{wang2021promises, huang2017gray, musavi2016experience}. The Model Context Protocol (MCP) addresses this by standardizing how AI models discover and interact with external tools and data sources through a client-server architecture in which MCP servers enable clients to access external information and perform actions such as calling APIs and executing scripts (\textit{tool-calling})~\cite{modelcontextprotocol_intro, modelcontextprotocol2025spec, hasan2025model}.
As MCP becomes embedded in AI applications, the reliability of AI models' interactions with MCP servers is critical. Erroneous behaviors and failures from MCP servers can spread to AI applications and users~\cite{gaire2025systematization, hasan2025model}. Despite the growth of the MCP ecosystem, the fault characteristics of MCP servers remain empirically unexplored. Existing studies examine MCP repository health, ecosystem structure, and security risks~\cite{hasan2025model, guo2025measurement, guo2025systematic}, but none characterize the runtime faults that arise during MCP execution.

This gap is important because MCP servers may be developed in different programming languages and with different frameworks, making language- or framework-specific fault models insufficient to characterize failures across the ecosystem. We therefore focus on runtime faults in MCP server implementations that affect MCP-facing functionality and are agnostic to the implementation language or framework. These failures concern MCP-facing runtime behavior across server implementations, including JSON-RPC message exchange, stateful connections, capability negotiation, tool exposure and execution, state and configuration management, progress tracking, cancellation, error reporting, and related integration behavior~\cite{modelcontextprotocol2025spec}.

An MCP server may compile successfully and conform to its specification while still exhibiting inadequate behavior due to improper tool invocation, mismanaged execution context, or protocol contract violations \cite{zhao2025mcp}. Such deviations may resemble omission, value, and timing failures studied in distributed systems~\cite{avizienis2004basic, cristian1991understanding, zhou2018fault}, yet manifest under additional constraints imposed by strict protocol enforcement and structured tool mediation. In this paper, we aim to construct a taxonomy of faults in MCP servers by analyzing GitHub repositories and surveying developers, both of which are common sources for taxonomy creation in software engineering~\cite{humbatova2020taxonomy,bensoussan2025taxonomy}. To reduce effort in manually analyzing repositories, we focus on the MCP servers, leaving the study of MCP clients, including agents and agentic systems, to future work.

A taxonomy can organize MCP runtime faults into explicit categories that developers and researchers can use for diagnosis, test design, fault injection, and reliability evaluation. For example, MCP fault categories can guide tests that check whether invalid tool schemas, missing message identifiers, or timeout violations are handled correctly. This use is consistent with prior work in other domains: a validated web fault taxonomy has been proposed for defining test cases that target specific fault classes and for building fault-seeding tools that inject artificial faults resembling real ones~\cite{marchetto2007empirical}; a taxonomy of real faults in deep learning systems has been used to assess how well existing mutation operators cover observed real faults and to motivate new mutation operators~\cite{humbatova2020taxonomy}; and mutation-testing research shows that artificially seeded faults are commonly used for test-suite evaluation, with mutant detection positively correlated with real-fault detection under controlled conditions~\cite{just2014mutants}.

Following a dependability taxonomy defining a fault as the cause of an error resulting in a deviation from service~\cite{avizienis2004basic}, we define an \textit{MCP server runtime fault} as a defect in an MCP server implementation that manifests during execution and disrupts MCP-facing functionality. Such faults may arise from protocol violations or from server-side implementation and integration mechanisms supporting MCP operation. An MCP server may remain operational while delivering incorrect service due to violations of the MCP session lifecycle, inconsistent declaration of server capabilities (\texttt{tools}, \texttt{resources}, or \texttt{prompts}) during handshake, schema mismatches in capability definitions, or incorrect propagation of tool results. Consistent with definitions of service failure in dependable systems~\cite{avizienis2004basic, cristian1991understanding}, such deviations manifest as omission, incorrect content, or timing violations at the service interface.

There are two approaches for creating a taxonomy: enumerative and faceted~\cite{meldrum2017crowdsourced,humbatova2020taxonomy}. Enumerative classification relies on predefined categories, whereas faceted classification derives categories inductively from empirical evidence. Because MCP represents a rapidly evolving ecosystem without established fault models, imposing predefined categories would risk introducing inappropriate abstractions. We therefore adopt a faceted, bottom-up construction process in which categories and subcategories emerge from observed evidence of failure. To derive the taxonomy, we manually analyzed 833 confirmed runtime fault threads identified through a repository-mining pipeline covering 473 actively maintained MCP server repositories on GitHub. Through iterative open coding and hierarchical consolidation, we organized recurring MCP faults into 11 top-level categories and 27 subcategories.

To assess whether the taxonomy reflects faults encountered in practice, we conducted an independent practitioner survey with 55 MCP server developers. We selected MCP server developers because they are directly involved in implementing, maintaining, and debugging the protocol-level behavior studied in this work, including capability declarations, tool schemas, session handling, and tool-result propagation. We used criterion-based purposive sampling, a non-probability sampling strategy in which participants are deliberately selected because they satisfy criteria relevant to the research objective~\cite{webster2023samplingmethods}. Specifically, we recruited maintainers and recent contributors from active MCP repositories in our dataset, since their public development activity provided evidence of relevant MCP server experience. This strategy is consistent with prior empirical studies that used purposive sampling to recruit practitioners with study-relevant expertise~\cite{ma2025practitioners,alami2021pull}. The survey was designed to validate the taxonomy’s practical relevance and coverage with developers who have direct MCP server experience.

Participants reported experiencing, on average, 20 of the 27 subcategories, and no category remained unobserved. Open-ended responses from participants did not reveal additional runtime fault classes. These findings indicate that the taxonomy captures a recognized coordination fault space within the MCP server ecosystem and provides a foundation for MCP reliability research, protocol-aware testing, and systematic fault injection.

\noindent\textbf{Contributions.}
The main contributions of this paper are:

\begin{itemize}
 
    \item The first empirical taxonomy of runtime faults in MCP servers, derived from the manual analysis of 833 confirmed runtime fault threads identified through a mining pipeline covering 473 actively maintained MCP server GitHub repositories.

    \item A bottom-up open coding methodology that organizes MCP runtime faults into 11 top-level categories, 27 subcategories, and 73 leaf fault types.

    \item An independent practitioner validation of the taxonomy through a survey of 55 MCP server developers, showing broad practical coverage, with no subcategory remaining entirely unobserved.

    \item A structured foundation for MCP reliability research, protocol-aware conformance testing, systematic fault injection, and runtime monitoring.
\end{itemize}

\textbf{Paper Organization.}
Section~\ref{sec:background} introduces MCP, and Section~\ref{sec:related} reviews prior work.  Section~\ref{sec:method} presents our methodology for creating and validating the taxonomy. Section~\ref{sec:result} presents the resulting taxonomy and reports its validation with practitioners.
Section~\ref{sec:discussion} discusses the implications of the findings, and Section~\ref{sec:threats} outlines threats to validity.
Section~\ref{sec:conclusion} concludes the paper.

\section{Background: Model Context Protocol}

\begin{figure}[!htbp]
    \centering
    \includegraphics[width=0.48\textwidth]{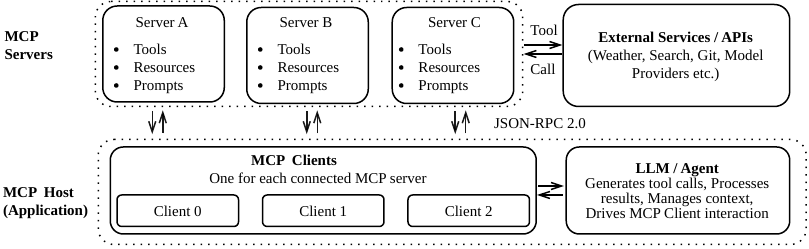}
    \caption{Overview of the Model Context Protocol (MCP) architecture.}
    \label{fig:mcp_architecture}
\end{figure}

Foundation models trained on broad data can be adapted to many downstream tasks~\cite{bommasani2021opportunities}, and are increasingly deployed in software systems for natural-language interaction, reasoning, and task automation. In practice, however, these applications often require access to external tools, data sources, and services beyond what is encoded in model parameters~\cite{yao2022react,schick2023toolformer}, a need partially addressed by structured function calling, where a model produces structured calls that the application executes~\cite{openai2025functioncalling}. The Model Context Protocol (MCP) generalizes this by providing an open standard that replaces per-application, system-specific integrations with a unified interface layer~\cite{modelcontextprotocol_intro,modelcontextprotocol2025spec}.
As shown in Figure~\ref{fig:mcp_architecture}, an LLM-enabled application operates as an \emph{MCP Host}, containing an LLM or agent that drives interaction and one \emph{MCP Client} instance per connected \emph{MCP Server}~\cite{modelcontextprotocol_intro,modelcontextprotocol2025spec}. Each server exposes three capability primitives, \emph{Tools} for callable operations, \emph{Resources} for contextual data, and \emph{Prompts} for reusable instruction templates~\cite{modelcontextprotocol2025spec}, and may invoke external services or APIs via tool calls. Communication is based on JSON-RPC~2.0~\cite{jsonrpc2013spec}, with the protocol layer defining message formats, capability negotiation, session lifecycle, cancellation, and error reporting, while the transport layer carries messages over mechanisms such as standard input/output or streamable HTTP~\cite{modelcontextprotocol2025spec}.
This architecture makes server correctness depend on coordinated runtime behavior across all of these protocol obligations. MCP servers may be implemented in different languages and frameworks, yet must satisfy the same interaction contract. Runtime faults therefore often arise at the boundary between protocol obligations and implementation behavior, and can remain invisible to build-time checks since a server may compile and start successfully while still violating MCP-specific runtime expectations during execution.

\label{sec:background}

\section{Related Work}
\label{sec:related}

\smallskip
\noindent\textbf{Classical Dependability Taxonomies.}
Avižienis et al.~\cite{avizienis2004basic} define failure as a deviation from specified service and distinguish omission, value, and timing violations. Cristian~\cite{cristian1991understanding} extends this model to distributed systems, identifying omission, response, timing, and crash failures at message interfaces. Chevochot and Puaut~\cite{chevochot2001experimental} show that distributed systems may continue execution while producing incorrect values or timing violations, and Huang et al.~\cite{huang2017gray} introduce the concept of gray failure, where degraded behavior remains undetected. These models formalize observable service deviations but abstract away protocol-specific interaction constraints. They do not characterize runtime fault mechanisms in schema-governed coordination systems such as MCP servers. Our work builds on these definitions while deriving an empirical taxonomy of MCP-specific runtime faults from developer evidence.

\smallskip
\noindent\textbf{Failures in API-driven Systems.}
A failure type in API-driven systems is API misuse, which violates usage constraints, such as incorrect call order, invalid parameters, and data format violations. These issues have been studied for library APIs~\cite{gu2019empirical}, web APIs~\cite{musavi2016experience}, and microservices~\cite{weng2018root}. Distributed systems also encounter interaction faults due to missing or incorrect coordination of components or services~\cite{weng2018root,zhou2018fault, zhang2024failure}. While prior work considers coordination faults in general API and microservice infrastructures, none provides an empirical taxonomy of runtime faults specific to MCP servers. MCP introduces schema validation, capability negotiation, and ordered tool execution, which shape the manifestation of failures yet remain unexplored.

\smallskip
\noindent\textbf{Mining Software Repositories for Failure Evidence.}
Repository artifacts provide empirical evidence of fault manifestation and resolution. Zhang et al.~\cite{zhang2018empirical} and Gao et al.~\cite{gao2025empirical} mined issue discussions to characterize bug symptoms and causes. Zhang et al.~\cite{zhang2018empirical} and Humbatova et al.~\cite{humbatova2020taxonomy} identified bugs and faults in deep learning systems using sources such as GitHub repositories and interviews with practitioners. Ning et al.~\cite{ning2026defining} derived an empirical defect taxonomy for LLM-based autonomous agents via large-scale repository mining. These works establish repository mining as a methodology for empirical fault analysis. However, none derive a bottom-up taxonomy of MCP-specific runtime coordination faults from repository discussions or validate it with practitioners. Our study addresses this gap.

\smallskip
\noindent\textbf{MCP Ecosystem and the Open Gap.} Recent studies have examined the structural and security properties of the MCP ecosystem. Guo et al.~\cite{guo2025measurement} measured ecosystem growth and adoption. Hasan et al.~\cite{hasan2025model} and Hou et al.~\cite{hou2025model} analyzed the health, architecture, and risk scenarios of MCP servers. Li et al.~\cite{li2025we} measured API privilege exposure, Zhao et al.~\cite{zhao2025mind} demonstrated toolchain attacks, and Watanabe et al.~\cite{watanabe2025use} studied fragility in agent-mediated tool interactions. These studies characterize ecosystem-scale architecture and security exposure but do not analyze developer-reported runtime faults in MCP servers. We aim to complement this MCP literature by systematically deriving and validating a taxonomy of MCP-specific runtime failures.

\section{Methodology}
\label{sec:method}

\begin{figure}[!htbp]
    \centering
    \includegraphics[width=0.49\textwidth]{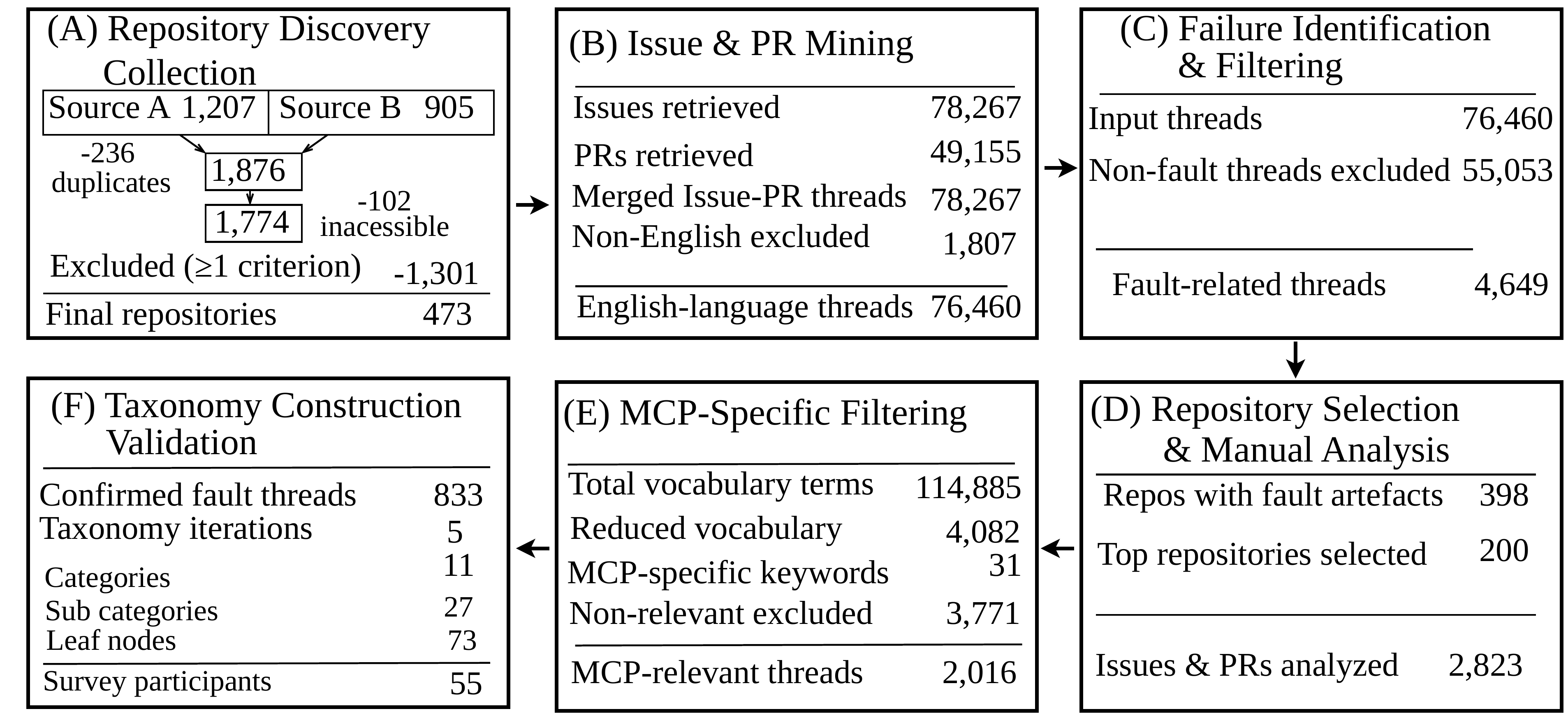}
    \caption{Overview of the methodological pipeline used in this study.}
    \label{fig:failure_modes}
\end{figure}

We adopted a bottom-up methodology inspired by prior studies on software fault taxonomies~\cite{humbatova2020taxonomy, thomas2024muprl, jahan2025taxonomy, nikanjam2022faults}. The methodology operationalizes the scope defined in the Introduction: we analyze server-side MCP runtime faults, not faults confined to MCP clients, host applications, agent orchestration logic, build configuration, deployment infrastructure, or general programming errors unrelated to MCP server functionality. During artifact filtering and manual labeling, a thread was retained only when the reported failure affected MCP-facing server functionality, including protocol-facing behavior such as JSON-RPC message exchange, capability exposure, connection handling, progress or cancellation handling, and structured error reporting~\cite{modelcontextprotocol2025spec}, as well as server-side implementation and integration behavior such as tool execution, state and configuration management, and provider integration.
Figure~\ref{fig:failure_modes} shows the main stages of the research design. The following subsections describe each stage in detail.

\subsection{Repository Discovery and Collection}

We compiled a comprehensive dataset of MCP tool–server repositories by combining two public sources: 
\textit{MCP Market}\footnote{\url{https://mcpmarket.com/server}}, a public registry indexing deployed MCP servers, and \textit{Awesome MCP Servers}\footnote{\url{https://github.com/punkpeye/awesome-mcp-servers}}, a community-maintained catalog of open-source implementations.
Repositories from MCP Market were collected using a custom crawler built with the BeautifulSoup library\footnote{\url{https://www.crummy.com/software/BeautifulSoup/bs4/doc/}}, which iterated through category and server pages to extract GitHub links. The Awesome MCP Servers list was parsed from its Markdown index using pattern-based regular expressions to identify repository URLs. All extracted entries were normalized to the canonical \texttt{owner/repo} format and cleaned to remove malformed URLs and cross-source duplicates. In total, the two sources yielded 2,112 entries (1,207 from MCP Market and 905 from Awesome MCP Servers). After removing 236 cross-source duplicates, the dataset contained 1,876 unique repository candidates. Each repository was validated through the GitHub REST API\footnote{\url{https://docs.github.com/en/rest/search?apiVersion=2022-11-28}} to ensure accessibility and metadata consistency. Repositories that returned HTTP~404 or were marked as private (102 in total) were excluded. 
The final dataset, therefore, contained 1,774 active, publicly accessible repositories. For each retained repository, we extracted metadata, including the number of commits, contributors, issues, pull requests, stars, and forks.

To focus on repositories representing real MCP systems actively used and maintained by developers, we applied exclusion criteria based on collaboration, activity, development history, popularity, and repository availability. These dimensions are commonly used in repository-mining studies to distinguish actively maintained engineered projects from personal, experimental, or abandoned repositories~\cite{kalliamvakou2016depth,munaiah2017curating, humbatova2020taxonomy,ustunboyacioglu2024data,owotogbe2025chaos}. Specifically, we excluded:
(i) personal repositories with fewer than five contributors, to reduce the influence of isolated or experimental projects;
(ii) inactive repositories, i.e., repositories with no open issues, to retain projects with observable maintenance activity;
(iii) repositories with trivial development history, defined as fewer than 100 commits, to ensure sufficient development evidence for mining fault-related artifacts;
(iv) unpopular repositories with fewer than 10 stars and fewer than 10 forks, to reduce noise from repositories with limited evidence of external use or community interest; and
(v) archived or disabled repositories, because they no longer represent actively maintained systems.

The reported counts are not mutually exclusive, as a repository may satisfy multiple criteria simultaneously. In total, 1,301 repositories were excluded, resulting in a final set of 473 MCP repositories that served as the basis for artifact mining and manual analysis. All exclusions were logged to support reproducibility. All repository data were collected in September 2025 and used as input to the repository mining pipeline.

\subsection{Issue and Pull Request Mining}

We mined issue and pull request data for 473 validated repositories using a custom crawler built on the GitHub REST API\footnote{\url{https://docs.github.com/en/rest}}, with token rotation and rate-limit handling integrated for stability\footnote{\url{https://docs.github.com/en/rest/rate-limit?apiVersion=2022-11-28}}. To balance coverage and feasibility, the crawler retrieved up to 200 of the most recent entries per repository, consistent with prior large-scale repository studies~\cite{khan2025integrating}. For each entry, we recorded its metadata (repository, issue number, URL, state, title, body, and closure timestamps), its labels, and whether it represented a pull request. Prior to assembling discussion threads, the mining stage produced 78,267 issues and 49,155 pull requests. Each artifact was stored as a structured message-level record containing author identifiers, timestamps, and associated text. To reconstruct the full conversational context, we linked pull requests to referenced issues using a regular expression matcher applied to the bodies of issues, comments, commit messages, and review text. The matcher captured GitHub’s standard reference patterns, including symbolic references (e.g., "\#123"), action phrases ("closes", "fixes", "resolves"), and explicit URLs. After merging each issue with its linked pull requests and associated artifacts, we obtained 78,267 consolidated issue-centered threads. To ensure linguistic consistency for downstream qualitative analysis, we applied automated language detection to the merged threads \cite{ehsani2025towards}. This step identified 77,619 English-language threads and 648 non-English threads, which were excluded. During subsequent manual inspection, we identified an additional 1,159 non-English threads that had not been detected automatically \cite{islam2025first}. After removing these, the final dataset comprised 76,460 English-language threads. At the time of data collection, 16,758 threads were open and 59,702 were closed; only the closed threads were carried forward to the fault-identification stage. Each merged thread corresponds to a single issue, optionally enriched with linked pull requests and related artifacts. The textual components were concatenated into a single \texttt{full\_thread\_text} field for analysis.
We analyzed full issue-centered threads rather than titles or labels alone because the evidence needed to distinguish confirmed runtime faults from feature requests, configuration questions, duplicates, or generic implementation bugs often appears in diagnostic comments, linked pull requests, commit messages, or closing discussions. When linked pull requests or commits are available, they also provide repair evidence that helps clarify the fault mechanism and its resolution.

\subsection{Failure Identification and Filtering}

Not all of the 59,702 closed threads (issue and pull request) describe faults relevant to MCP systems. To identify fault-fixing activities, we applied a rule-based filtering step inspired by prior studies on fault taxonomies~\cite{humbatova2020taxonomy, morovati2023bugs}. For issue and pull request threads, we retained artifacts labeled as bug, defect, or error. For commits, we mined commit messages containing the terms ("fix" or "solve") and ("bug", "issue", "problem", "defect", or "error"). The thread-level filter retained 4,649 closed threads and excluded the remaining 55,053; 398 repositories contained at least one retained fault-related thread.

\subsection{Selection of Repositories for Manual Analysis}

Following prior work on empirical fault taxonomies~\cite{humbatova2020taxonomy}, we selected a subset of repositories for manual analysis. We used criterion-based repository selection rather than random sampling because the objective was to maximize the diversity and recurrence of observable MCP runtime fault patterns, not to estimate repository-level fault prevalence. Specifically, we selected the 200 repositories with the highest number of fault-related issues and pull requests among the repositories that passed the fault-related filter. This information-rich selection strategy is appropriate when the goal is qualitative coverage of relevant cases rather than statistical generalization~\cite{baltes2022sampling}. Repositories with more fault-related discussions are more likely to contain richer evidence of recurring failures, developer diagnosis, and fixes. This choice made the manual qualitative analysis tractable while preserving a broad set of fault manifestations. The selected repositories contained 2,823 issues and pull requests, and 937 commits associated with fault-fixing activities. Thus, the 473 repositories constitute the initial mining pool, whereas the 200 repositories constitute the sample selected for detailed MCP-specific filtering and manual analysis.

\subsection{MCP-Specific Relevance Filtering}

Before the manual labeling, we inspected a random sample of 100 candidate artifacts and found a high proportion of false positives (80\%), corresponding to generic programming issues unrelated to MCP-specific mechanisms. This additional filtering was necessary because generic bug labels and fault-related keywords frequently captured ordinary implementation problems, whereas our study targets runtime faults affecting MCP server functionality. To reduce noise and focus the analysis on MCP-relevant faults, we applied an additional relevance-filtering step inspired by prior empirical fault-taxonomy work~\cite{humbatova2020taxonomy}.

We extracted the vocabulary of stemmed words from the mined issues, pull requests, and commits, yielding 114,885 unique terms. For commits, terms were extracted from commit messages, while for issues and pull requests, we considered titles, descriptions, and all discussion comments. We sorted terms by frequency and removed the long tail of words appearing in fewer than 10 artifacts, yielding a reduced vocabulary of 4,082 terms. These terms were manually inspected by the first two authors to identify MCP-relevant terminology related to tool invocation, schema usage, context management, and LLM-mediated interaction. All flagged terms were discussed in a joint meeting, resulting in a final vocabulary of 31 MCP-specific keywords. After excluding artifacts that did not contain these keywords, we obtained 2,016 candidate issue-centered threads spanning 177 of the 200 selected repositories. Thus, the 177 repositories represent those contributing MCP-relevant candidate threads at this filtering stage. Each of the 2,016 threads was then manually inspected to determine
whether it described a runtime fault affecting MCP server functionality. During this inspection, 71 non-English threads were excluded. Among the remaining threads, 623 did not correspond to fault-fixing activities, and 489 described build- or environment-related errors outside the scope of MCP runtime faults. The final dataset comprised 833 confirmed runtime fault threads used for taxonomy construction.

\subsection{Taxonomy Construction and Validation}

\smallskip
\noindent\textbf{Taxonomy Construction.}
We constructed the MCP runtime fault taxonomy using a bottom-up qualitative approach inspired by prior empirical fault taxonomy studies~\cite{humbatova2020taxonomy, vijayaraghavan2003bug}. The analysis was grounded in 833 confirmed MCP-specific runtime fault threads. Throughout this process, we also referred to the official MCP documentation~\cite{modelcontextprotocol_intro} to ensure the accuracy of terminology and conceptual understanding. The taxonomy was developed through five iterative refinement rounds. All artifacts were coded independently by the first and second authors, with disagreements resolved through discussion and consensus. In \emph{Round 1}, all 833 threads were open-coded using fine-grained descriptive fault labels. In \emph{Round 2}, semantically equivalent labels were normalized and merged. In \emph{Round 3}, the normalized labels were organized using "is-a" relationships, yielding 15 provisional top-level categories and 44 subcategories. In \emph{Round 4}, overlapping categories were merged and terminology aligned with MCP constructs, resulting in 11 top-level categories and 27 subcategories. In \emph{Round 5}, all threads were reclassified using the finalized codebook; at the final subcategory level, 52 disagreements occurred and Cohen's $\kappa=0.938$, with disagreements resolved by consensus. No new structural categories emerged, indicating saturation. The final taxonomy comprises 11 top-level categories, 27 subcategories, and 73 leaf fault types.

\smallskip
\noindent\textbf{Survey Design.}
To evaluate whether the taxonomy reflects runtime faults encountered in practice, we conducted an independent practitioner survey, drawing on prior studies that have validated software engineering taxonomies~\cite{humbatova2020taxonomy}. The survey was administered via Google Forms and comprised three sections. The first section collected demographic information, including job role and years of professional experience in software engineering and MCP-related development. The second section evaluated the provisional taxonomy available at the time of survey administration, comprising 45 fault items. Each item was presented with a description and representative examples. Subsequent taxonomy refinement consolidated these items into the 27 subcategories of the final taxonomy; the complete item-to-subcategory mapping is provided in the replication package.

Participants indicated whether they had encountered the fault in practice. Severity and diagnostic effort were collected to assess the practical salience of each category beyond simple recognition. Severity ratings indicate the perceived impact, whereas effort ratings reflect the cognitive and engineering complexity of diagnosing and resolving the fault. Together with open-ended responses that invite participants to report missing runtime fault categories, these measures provide complementary validation of both the structural completeness and the practical relevance of the taxonomy. For confirmed categories, respondents rated severity (Minor, Major, Critical) and the effort required to identify and diagnose the issue (Low, Medium, High). The third section included open-ended questions that invited participants to report additional MCP runtime faults not covered by the taxonomy and to provide feedback on the taxonomy's clarity and usefulness.

\smallskip
\noindent\textbf{Validation Procedure and Participants.}
We selected MCP server practitioners because they are directly involved in implementing, maintaining, and debugging the MCP-facing server functionality studied in this work, including capability declarations, tool schemas, session handling, and tool-result propagation. We used criterion-based purposive sampling, a non-probability sampling strategy in which participants are deliberately selected because they satisfy criteria relevant to the research objective~\cite{baltes2022sampling,webster2023samplingmethods}. Specifically, approximately 1,200 maintainers and recent contributors from active MCP repositories in our dataset were invited to participate via GitHub. Of these, 55 respondents completed the survey, corresponding to an approximate response rate of 4.6\%. This recruitment strategy is consistent with prior empirical studies that used purposive sampling to recruit practitioners with study-relevant expertise~\cite{ma2025practitioners,alami2021pull}. None of the respondents was involved in taxonomy construction, preserving independence between taxonomy derivation and practitioner validation.

Participants included software engineers, architects, AI/ML engineers, DevOps engineers, technical leads, and founders. Overall professional experience ranged from less than 1 year to more than 10 years, with the largest group (21/55) reporting over 10 years of experience, and the median fell within the 5-10-year range. Experience with MCP servers or related LLM tool-integration systems ranged from less than 1 year to over 3 years, with most respondents (24/55) reporting 1-2 years of relevant development experience. Survey responses were analyzed to determine (i) the proportion of fault categories encountered in practice, (ii) perceived severity distributions, and (iii) reported diagnostic effort. Open-ended responses were reviewed qualitatively. No additional MCP-specific runtime fault category emerged, supporting the taxonomy's empirical coverage and stability.

\section{Results: MCP Server Fault Taxonomy}
\label{sec:result}

\begin{figure*}[p]
    \centering
    \includegraphics[
        angle=90,
        origin=c,
        width=0.65\textheight,
        keepaspectratio
    ]{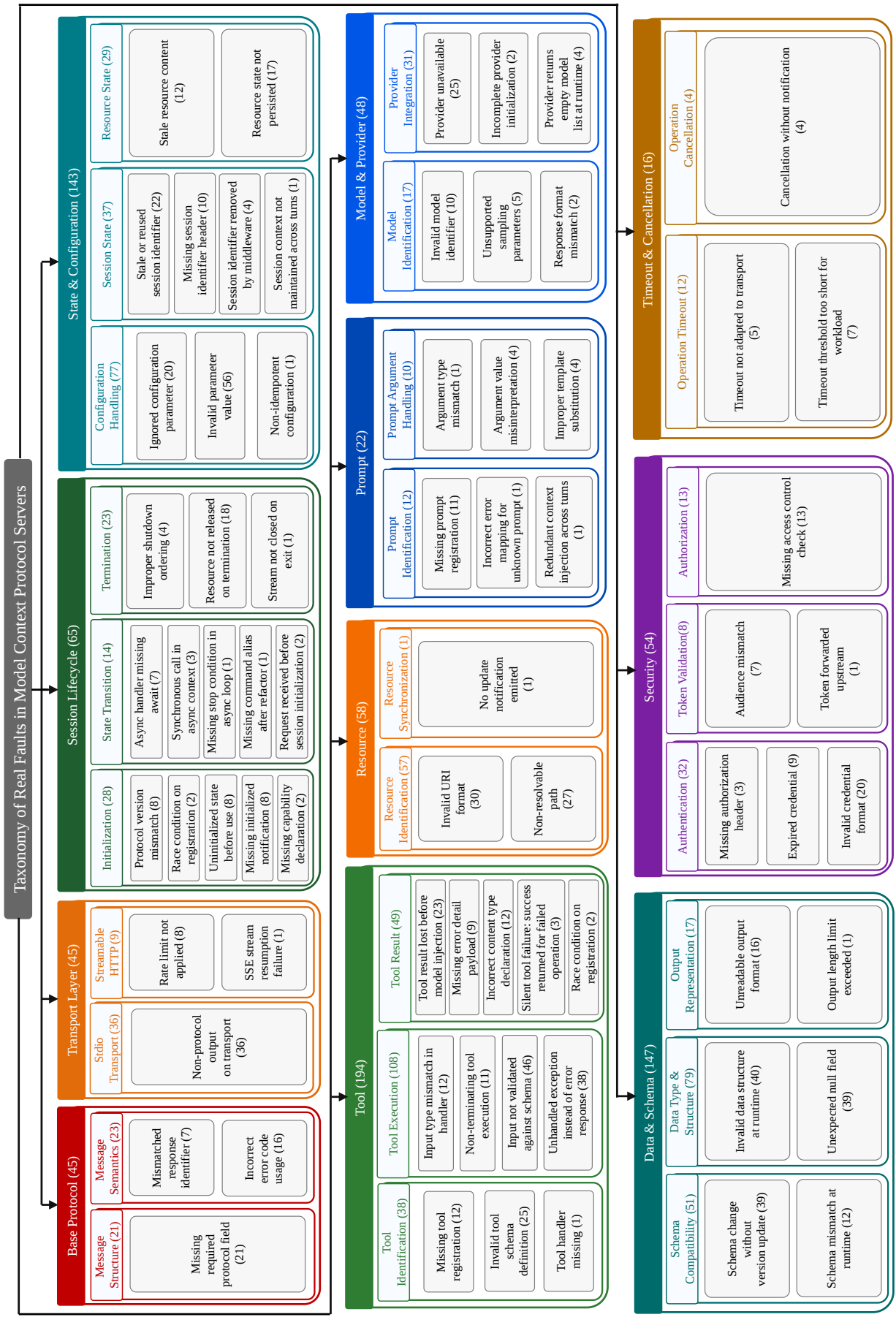}
    \caption{Taxonomy of MCP Server Failure Modes.}
    \label{fig:cm_mode}
\end{figure*}

Figure~\ref{fig:cm_mode} shows the final runtime fault taxonomy for MCP servers, comprising 11 top-level categories with corresponding subcategories. The parenthetical numbers indicate the count of confirmed fault threads assigned to each category during manual analysis. The remainder of this section discusses each category in detail.

\subsection{Base Protocol}

This category of the taxonomy covers faults that affect the correctness of JSON-RPC 2.0 interactions underlying message exchange in MCP. It captures violations of JSON-RPC 2.0 request–response structure, request–response correlation via the \texttt{id} field, and error object semantics.

\emph{Message Structure}. This considers faults that affect conformance with the JSON-RPC 2.0 request and response object structure. It includes omission or improper propagation of required fields such as \texttt{id}, \texttt{method}, \texttt{params}, \texttt{result}, or \texttt{error}. These faults concern message validity rather than errors in session initialization or capability negotiation. For example, a thread in the \texttt{arize-ai/openinference}\footnote{\url{https://github.com/Arize-ai/openinference/issues/2240}} repository reports that a tool execution response returned an empty \texttt{id} field. The issue notes that the \texttt{tool\_call id} was missing from the payload, preventing correct correlation between the request and its result. The fix ensured that identifiers were propagated in the response.

\emph{Message Semantics}. Faults in this category affect the semantic interpretation of otherwise well-formed JSON-RPC 2.0 messages. They arise when identifiers are mismatched across related events or when application-level failures are incorrectly represented as HTTP status codes rather than structured JSON-RPC \texttt{error} objects. Unlike structural faults, the message format conforms to JSON-RPC 2.0, but its role within the protocol interaction is incorrect. A representative instance appears in the \texttt{CherryHQ/cherry-studio}\footnote{\url{https://github.com/CherryHQ/cherry-studio/pull/10781}} repository, where a streaming \texttt{text-end} event referenced an incorrect identifier, preventing correlation with its corresponding start event. The fix enforced consistent identifier usage across related events, restoring the correct cross-event identifier linkage.

\subsection{Transport Layer}

This category covers faults related to the mechanisms used to transport MCP messages between the client and the server. MCP defines transport bindings, including stdio for local process communication and Streamable HTTP for remote interaction. Faults in this category arise when transport behavior violates framing, message boundary, or session-handling requirements defined by the MCP transport binding.

\emph{Stdio Transport}. This subcategory considers faults specific to the stdio transport, where MCP messages are exchanged over standard input and output streams. MCP requires that only valid JSON-RPC 2.0 messages be emitted on standard output. Faults arise when implementations write non-protocol data to stdout, thereby corrupting the message stream and breaking client parsing. For example, a thread in the \texttt{circleci-public/mcp-server-circleci}\footnote{\url{https://github.com/CircleCI-Public/mcp-server-circleci/issues/125}} repository reports that the MCP server printed non-MCP command output to standard output. As described in the issue, this unexpected output interfered with proper message processing over the stdio channel.

\emph{Streamable HTTP}. Faults in this subcategory affect MCP deployments using the Streamable HTTP transport binding. These issues concern incorrect handling of HTTP requests, session identifiers, or server-sent event streams under the MCP transport specification. For example, a thread in the \texttt{awslabs/mcp}\footnote{\url{https://github.com/awslabs/mcp/issues/1377}} repository reports that the server did not apply request throttling and was unaware of excessive requests. The issue highlights the absence of operational safeguards in an HTTP-based MCP deployment, exposing weaknesses in transport-layer management under high-volume traffic.

\subsection{Session Lifecycle}

This category of the taxonomy covers faults arising from violations of the initialization handshake, capability negotiation, session state progression, and termination semantics as defined by MCP. These faults arise when servers or clients fail to implement the lifecycle requirements for a valid MCP session correctly.

\emph{Initialization}. This subcategory captures faults occurring during the setup phase of an MCP session, before the system reaches a fully initialized state. It includes protocol version mismatches during the initialization handshake, race conditions during capability negotiation or registration, use of uninitialized state, and missing initialized notifications. Such faults prevent the session from reaching a consistent ready state. For example, a thread in the \texttt{cgoinglove/better-chatbot}\footnote{\url{https://github.com/cgoinglove/better-chatbot/pull/220}} repository reports an "MCP infinite loading issue" caused by incorrect MCP manager initialization. The associated commit notes the removal of a redundant initialization call and the introduction of a \texttt{waitInitialized} mechanism, indicating that the system was operating before proper initialization had completed.

\emph{State Transition}. Faults in this subcategory affect the correct progression of an active MCP session after initialization. These include async handlers missing \texttt{await}, synchronous calls executed in an asynchronous context, missing stop conditions in async loops, requests received before session initialization, and missing command aliases after refactoring. Such faults do not concern message structure, but rather the correctness of request handling and control flow within an active MCP session. A case appears in the \texttt{jlowin/fastmcp}\footnote{\url{https://github.com/jlowin/fastmcp/pull/2084}} repository, where the CLI invoked a synchronous \texttt{server.run()} method from within an already running async context, raising an "Already running asyncio in this thread" error. The fix replaced the synchronous calls with their async equivalents, restoring correct control flow during an active session.

\emph{Termination}. This subcategory captures faults occurring during the shutdown of an MCP session. These include improper shutdown ordering, failure to release resources on termination, and input streams not being closed on exit. Such faults may leave the server in an inconsistent state or cause runtime errors during subsequent sessions. For example, a thread in the \texttt{BeehiveInnovations/zen-mcp-server}\footnote{\url{https://github.com/BeehiveInnovations/zen-mcp-server/issues/181}} repository reports improper shutdown ordering during server termination. The cleanup logic was executed out of order, necessitating restructuring the shutdown sequence to ensure that resources were released correctly before process exit.

\subsection{State and Configuration}

This category covers faults in implementation-level state management and configuration handling within MCP deployments, including server-level configuration parameters, session identifiers, and resource persistence. These faults do not concern JSON-RPC message structure or lifecycle semantics, but rather the correctness and consistency of state management across requests.

\emph{Configuration Handling}. This subcategory captures faults arising from improper interpretation or enforcement of server-level configuration parameters. It includes ignored configuration flags and invalid parameter values that result in behavior inconsistent with the documented semantics. For example, a thread in the \texttt{apache/apisix}\footnote{\url{https://github.com/apache/apisix/issues/12537}} repository reports that setting \texttt{include\_resp\_body} to \texttt{false} in the \texttt{http-logger} plugin did not suppress response bodies from log output. The investigation revealed that a second plugin, \texttt{file-logger}, had the same flag set to \texttt{true}, and because both plugins share the same logging utility, the \texttt{true} setting overrode the \texttt{false} one, causing the configuration parameter to be silently ignored.

\emph{Session State}. Faults in this subcategory affect the correctness of session identifiers, such as session identifier headers used to correlate requests within an MCP session. These include stale or reused session identifiers, absent session identifier headers, and identifiers removed by middleware layers. Such faults break session continuity without violating JSON-RPC structure. A representative example appears in the \texttt{arize-ai/openinference}\footnote{\url{https://github.com/Arize-ai/openinference/issues/2313}} repository, where a fix addresses "session ID issues" and improves \texttt{session\_id} extraction in streaming paths. The discussion indicates that incorrect or missing session identifiers led to inconsistent request correlation across streaming interactions.

\emph{Resource State}. This subcategory captures faults related to the persistence and freshness of server-managed resource state exposed through MCP resource operations. These include stale resource content and failure to persist state changes across operations. For example, a thread in the \texttt{bucketco/bucket-javascript-sdk}\footnote{\url{https://github.com/bucketco/bucket-javascript-sdk/pull/286}} repository reports a bug where overrides did not appear in the API result, prompting a fix to "propagate updates". The issue arises from a failure to persist or expose the updated resource state, resulting in clients observing outdated content.

\subsection{Tool}

This category of the taxonomy covers faults related to the declaration, invocation, and result propagation of tools within MCP. These faults arise when tool capabilities, execution handlers, or returned outputs do not conform to the tool capability declaration and invocation rules defined by MCP, independently of transport or session lifecycle correctness.

\emph{Tool Identification}. This subcategory captures faults affecting the declaration of tools within the server’s capabilities object and their associated input/output schemas. It includes a missing tool capability declaration, where expected tools are not exposed through capability negotiation, and invalid schema specifications that violate the declared input or output structure. Such faults prevent tools from being discoverable or callable, even with correct transport and session state. For example, a thread in the \texttt{stripe/agent-toolkit}\footnote{\url{https://github.com/stripe/agent-toolkit/issues/86}} repository reports that dispute-related tools were "not available on MCP," indicating that the expected tools were not properly declared or exposed through the server’s capabilities.

\emph{Tool Execution}. Faults here affect tool handlers after the capability declaration. These include input-type mismatches in handlers, non-terminating execution, failures to validate inputs, and runtime errors not represented as JSON-RPC \texttt{error} objects. Such faults occur in tool invocation rather than identification. An example appears in the \texttt{alfonsograziano/node-code-sandbox-mcp}\footnote{\url{https://github.com/alfonsograziano/node-code-sandbox-mcp/issues/42}} repository, where tool execution entered a non-terminating state, preventing completion of the request and blocking the MCP interaction loop.

\emph{Tool Result Propagation}. This subcategory captures faults that occur after tool execution completes but before the result is correctly returned in the JSON-RPC response and incorporated into the client or model context. These include missing result payloads, incorrect content structure, and cases where a success response is returned despite an underlying tool failure. Such faults do not concern the tool’s internal execution logic, but the correctness of result handling under MCP. For example, a thread in the \texttt{8b-is/smart-tree}\footnote{\url{https://github.com/8b-is/smart-tree/issues/21}} repository reports that tool results were not properly incorporated back into the execution context, leading to incomplete responses despite successful tool invocation.

\subsection{Resource}
MCP represents resources using URI-based identifiers and specifies operations, including \texttt{resources/list}, \texttt{resources/read}, and \texttt{resources/updated} notifications. This category includes faults related to protocol-level resource identification and synchronization. 

\emph{Resource Identification}. This subcategory captures faults that affect the correctness of resource URI formats and server-side resolution. It includes invalid URI structures and non-resolvable resource paths, preventing clients from successfully invoking \texttt{resources/read} or related operations. These faults concern resource addressing under MCP rather than implementation-level persistence or session state. For example, a thread in the \texttt{basicmachines-co/basic-memory}\footnote{\url{https://github.com/basicmachines-co/basic-memory/issues/304}} repository reports that the \texttt{build\_context} function failed to resolve \texttt{memory://} resource URIs containing underscore-formatted relation segments. While hyphenated paths such as \texttt{part-of/} returned results, underscore-formatted paths such as \texttt{part\_of/} silently returned empty results despite matching content existing in the underlying markdown files. The root cause was a mismatch between the URI format accepted as input and the normalized format used internally by the MCP server, resulting in a resource identification failure.

\emph{Resource Synchronization}. This subcategory captures faults in which changes to server-managed resource state are not correctly communicated to clients. These faults occur when the underlying resource state or operation remains valid, but its latest status is not reflected externally. Unlike \emph{Resource State}, which concerns stale or unpersisted resource data, Resource Synchronization concerns the propagation of resource changes. MCP supports such synchronization through \texttt{resources/subscribe} and \texttt{notifications/resources/updated} when resource subscriptions are enabled~\cite{modelcontextprotocol2025spec}. For example, a thread in the \texttt{coleam00/Archon}\footnote{\url{https://github.com/coleam00/Archon/issues/589}} repository reports that a knowledge-base crawl continued running in the background while its progress disappeared from the user interface. The crawl remained active, but its current status was no longer reflected to the client-facing interface, resulting in a resource synchronization failure.

\subsection{Prompt}

This category of the taxonomy covers faults related to the declaration and invocation of prompt capabilities within MCP. These faults arise when prompts are not correctly declared in the server’s capabilities object or when supplied arguments violate the declared input schema, independently of transport or session lifecycle correctness.

\emph{Prompt Identification}. This subcategory captures faults affecting the correct identification and resolution of prompts under MCP semantics. These include failures in prompt lookup, incorrect handling of unknown prompt names, or improper mapping of prompt-related errors. For example, a thread in the \texttt{jlowin/fastmcp}\footnote{\url{https://github.com/jlowin/fastmcp/issues/2127}}
 repository reports that fetching a non-existing prompt raised a \texttt{NotFoundError}, but the middleware incorrectly transformed it into an MCP Internal Error (\texttt{-32603}) instead of the appropriate MCP Resource Not Found error (\texttt{-32001}). 

\emph{Prompt Argument Handling}. Faults in this subcategory affect the correctness of arguments supplied during prompt invocation. These include violations of the declared input schema, argument-type mismatches, and incorrect value substitution, where runtime values do not conform to the declared JSON Schema types. Such faults occur after prompt discovery but before successful prompt execution. A representative example appears in the \texttt{arize-ai/openinference}\footnote{\url{https://github.com/arize-ai/openinference/issues/2269}} repository, where validation failed with the message "Invalid type dict in attribute 'llm.input\_messages.0.message.content' value sequence," indicating that a dictionary was passed where a different JSON Schema type was expected.

\subsection{Model Provider Integration}

This category covers faults arising at the boundary between an MCP server and its configured model provider. MCP servers expose model capabilities that depend on an underlying provider implementation. Correct operation depends on valid model selection, supported sampling parameters, and proper initialization of the provider. Faults in this category arise when provider configuration or invocation behavior does not conform to the capabilities supported by the underlying provider.

\emph{Model Identification}. This subcategory captures faults related to incorrect model selection and sampling parameter specification. It includes invalid model identifiers, unsupported sampling parameters, and attempts to invoke models not available through the configured provider. These faults do not concern MCP transport or lifecycle semantics, but rather the correctness of provider-backed model invocation within an MCP session. For example, a thread in the \texttt{modelcontextprotocol/inspector}\footnote{\url{https://github.com/modelcontextprotocol/inspector/issues/451}} repository reports a failure caused by specifying a model identifier not recognized by the configured provider, causing the server to attempt to invoke a nonexistent model.

\emph{Provider Integration}. Faults here affect the availability and readiness of the configured model provider. These include provider unavailability, incomplete provider initialization, and failure to list available models during capability negotiation. Such faults prevent the MCP server from correctly exposing or invoking models, even when protocol interactions are valid. A representative case appears in the \texttt{awslabs/mcp}\footnote{\url{https://github.com/awslabs/mcp/issues/1377}}
repository, where the provider returned no available models during runtime, due to incomplete or failed provider initialization.

\subsection{Data and Schema}

This category covers faults arising from inconsistencies between JSON Schema definitions declared in MCP capability specifications and the structured payloads used at runtime.

\emph{Schema Compatibility}. This captures faults when emitted content does not conform to the JSON Schema for a capability. These include missing properties or structural deviations that violate the output schema. For example, a thread in the \texttt{basicmachines-co/basic-memory}\footnote{\url{https://github.com/basicmachines-co/basic-memory/issues/263}} repository reports that invoking \texttt{build\_context} failed with the error "Structured content does not match the tool's output schema." The discussion identifies missing required properties and a violation of the \texttt{date-time} format in \texttt{created\_at}, indicating that the payload did not satisfy the declared JSON Schema.

\emph{Data Type and Structure}. Faults here arise when runtime payloads conform to the declared schema but contain values that are valid under the interface yet violate downstream implementation assumptions. These include permitted \texttt{null} values, partially populated objects, or context-dependent semantic inconsistencies. A representative example appears in the \texttt{apache/apisix}\footnote{\url{https://github.com/apache/apisix/issues/12548}} repository, where the \texttt{content} field returned by an LLM service was null. Although the interface accepted this value, downstream logic failed due to an implicit assumption that the string was non-null.

\emph{Output Representation}. This subcategory captures faults related to improper serialization or structured output formatting that, while syntactically valid JSON, do not conform to the declared structured output representation. These include malformed object representations that hinder correct interpretation despite a valid JSON-RPC structure. For example, a thread in the \texttt{datalayer/jupyter-mcp-server}\footnote{\url{https://github.com/datalayer/jupyter-mcp-server/issues/92}} repository reports that a \texttt{read\_cell} call returned source code as one-character-per-line arrays. Although the response was valid JSON-RPC, the structured content representation deviated from the expected object format.

\subsection{Security}

This category covers faults related to authentication, token validation, and authorization in MCP deployments. MCP servers commonly operate behind OAuth 2.1–protected HTTP endpoints and rely on bearer tokens for client authentication. Faults in this category arise when identity verification or access control mechanisms are missing, misconfigured, or inconsistently enforced.

\emph{Authentication}. This subcategory captures faults that affect the presence and format of credentials during request processing. It includes missing \texttt{Authorization} headers, expired bearer tokens, and malformed token values. These faults prevent proper client authentication before JSON-RPC requests are processed. For example, a thread in the \texttt{awslabs/mcp}\footnote{\url{https://github.com/awslabs/mcp/issues/1377}} repository reports request failures caused by absent or improperly propagated authorization headers, resulting in rejected MCP interactions due to missing authentication data.

\emph{Token Validation}. Faults in this subcategory concern incorrect verification of authentication tokens upon receipt. These include mismatches in the token audience (\texttt{aud}) claim during validation and cases where tokens are forwarded upstream without proper verification. Unlike authentication faults, the credential may be present but is not correctly validated against expected claims. A representative instance appears in the \texttt{BeehiveInnovations/zen-mcp-server}\footnote{\url{https://github.com/BeehiveInnovations/zen-mcp-server/issues/181}} repository, where token validation failed due to a claim mismatch, preventing successful request handling. 

\emph{Authorization}. This captures faults involving missing access control checks after authentication. These faults occur when MCP capabilities or tool invocations are executed without verifying client permissions. For example, a thread in the \texttt{basicmachines-co/basic-memory}\footnote{\url{https://github.com/basicmachines-co/basic-memory/pull/215}} repository discusses execution paths that lacked access control enforcement, demonstrating a missing authorization check in tool invocation.

\subsection{Timeout and Cancellation}

This category includes faults caused by premature termination or improper interruption of operations during an active MCP session. They generally arise when implementation-level timeout handling or cancellation request processing is misaligned with workload characteristics or stream termination, leading to incomplete or inconsistent request handling.

\emph{Operation Timeout}. This subcategory captures faults, where operations exceed predefined execution limits or timeout thresholds are not properly adapted to task complexity or streaming behavior. These faults occur at the implementation level rather than being mandated by MCP itself. Typical cases include timeout thresholds that are too short for long-running tool calls or streaming operations. For example, a thread in the \texttt{browser-use/browser-use}\footnote{\url{https://github.com/browser-use/browser-use/issues/3214}} repository reports a failure after 15 seconds with the warning "handler ... has been running for 15s on event," followed by a timeout error during text input processing. The discussion indicates that the fixed 15-second threshold was insufficient for the workload, resulting in the operation being terminated prematurely.

\emph{Operation Cancellation}. Faults in this subcategory affect cancellation handling and stream-termination signaling. These include cancellation requests that fail to propagate through the session and stream-termination requests ignored by execution handlers. Such faults may leave the tool execution or a JSON-RPC response stream in an unresolved state. For example, a thread in the \texttt{langchain-ai/langgraphjs}\footnote{\url{https://github.com/langchain-ai/langgraphjs/issues/1610}} repository reports that invoking \texttt{stream.stop()} failed to terminate the stream, causing incorrect cancellation during streaming operations.

\begin{table}[t]
\centering
\scriptsize
\caption{MCP Fault Taxonomy Validation Results ($n=55$)}
\label{tab:mcp_validation}
\resizebox{\linewidth}{!}{
\begin{tabular}{lcccccccc}
\hline
\textbf{Sub-Category} &
\multicolumn{2}{c}{\textbf{Response}} &
\multicolumn{3}{c}{\textbf{Severity}} &
\multicolumn{3}{c}{\textbf{Effort Required}} \\
\cline{2-3} \cline{4-6} \cline{7-9}
& Yes\% & No\% & Minor & Major & Critical & Low & Medium & High \\
\hline

Message Structure & 60 & 40 & 61 & 21 & 18 & 52 & 33 & 15 \\
Message Semantics & 49 & 51 & 44 & 37 & 19 & 22 & 59 & 19 \\

stdio Transport & 47 & 53 & 42 & 50 & 8 & 31 & 54 & 15 \\
Streamable HTTP & 36 & 64 & 39 & 33 & 28 & 26 & 53 & 21 \\

Initialization & 36 & 64 & 75 & 15 & 10 & 45 & 30 & 25 \\
State Transition & 47 & 53 & 23 & 62 & 15 & 27 & 38 & 35 \\
Termination & 25 & 75 & 29 & 50 & 21 & 29 & 50 & 21 \\

Configuration Handling & 33 & 67 & 61 & 28 & 11 & 50 & 33 & 17 \\
Session State & 36 & 64 & 50 & 40 & 10 & 40 & 35 & 25 \\
Resource State & 31 & 69 & 53 & 35 & 12 & 38 & 56 & 6 \\

Tool Identification & 42 & 58 & 57 & 35 & 9 & 52 & 43 & 4 \\
Tool Execution & 73 & 27 & 42 & 52 & 5 & 35 & 55 & 10 \\
Tool Result Propagation & 67 & 33 & 44 & 47 & 8 & 47 & 33 & 19 \\

Resource Identification & 25 & 75 & 64 & 14 & 21 & 62 & 23 & 15 \\
Resource Synchronization & 29 & 71 & 44 & 31 & 25 & 31 & 44 & 25 \\

Prompt Identification & 47 & 53 & 62 & 19 & 19 & 65 & 19 & 15 \\
Prompt Argument Handling & 36 & 64 & 47 & 32 & 21 & 40 & 40 & 20 \\

Model Identification & 25 & 75 & 50 & 29 & 21 & 50 & 29 & 21 \\
Provider Integration & 51 & 49 & 32 & 54 & 14 & 36 & 43 & 21 \\

Schema Compatibility & 58 & 42 & 53 & 40 & 7 & 60 & 33 & 7 \\
Data Type \& Structure & 56 & 44 & 65 & 23 & 13 & 61 & 26 & 13 \\
Output Representation & 55 & 45 & 60 & 30 & 10 & 63 & 20 & 17 \\

Authentication & 55 & 45 & 47 & 40 & 13 & 41 & 38 & 21 \\
Token Validation & 42 & 58 & 48 & 43 & 9 & 36 & 32 & 32 \\
Authorization & 40 & 60 & 27 & 55 & 18 & 24 & 43 & 33 \\

Operation Timeout & 44 & 56 & 42 & 42 & 17 & 38 & 42 & 21 \\
Operation Cancellation & 29 & 71 & 44 & 50 & 6 & 62 & 19 & 19 \\

\hline
\end{tabular}}
\end{table}

\subsection{Validation Results}
Table~\ref{tab:mcp_validation} summarizes the feedback from the practitioners ($n=55$). All 27 sub-categories were reported by at least 25\% of respondents, with an average confirmation rate of 44\%, indicating that each fault class reflects issues encountered in practice. The most frequently reported categories are \emph{Tool Execution} (73\%), \emph{Tool Result Propagation} (67\%), and \emph{Message Structure} (60\%), showing that faults related to tool execution, result propagation, and protocol message structure are recognized in practice. The least-reported categories, \emph{Resource Identification}, \emph{Termination}, and \emph{Model Identification} (25\%), were nevertheless identified by a minority, while \emph{Resource Synchronization} was reported by 29\% of respondents. Severity and effort ratings further distinguish impact. \emph{State Transition} faults stand out, with 77\% of respondents rating them Major or Critical and 35\% reporting High diagnostic effort, indicating substantial perceived impact and diagnostic difficulty. \emph{Authorization} also shows elevated impact, with 55\% rated as Major and 33\% reporting High effort. In contrast, categories such as \emph{Initialization} and \emph{Configuration Handling} were more frequently rated as Minor. Suggestions regarding ecosystem maturity, SDK instability, or deployment-level concerns fell outside the scope of runtime MCP server faults considered in this paper. No new runtime fault category emerged, supporting the taxonomy’s structural coverage and stability.

\section{Discussion}
\label{sec:discussion}

\smallskip
\noindent\textbf{Discussion of Results.}
This study presents the first empirical taxonomy of runtime faults in MCP-based systems, derived from 833 fault threads across 473 MCP server GitHub repositories and validated by 55 practitioners. 


Tool faults are the most frequent (192/833, 23.0\%), followed by Data~\&~Schema (147/833, 17.6\%) and State~\&~Configuration (143/833, 17.1\%). Together, these three categories account for more than half of all confirmed cases (482/833, 57.9\%). This distribution indicates that the dominant failure surface lies in capability execution and data schema enforcement, rather than in the transport or protocol layers. This pattern is consistent with prior studies of API-driven systems, where parameter misuse, configuration errors, and data-format violations are recurring root causes \cite{gu2019empirical,musavi2016experience}. 

We also observe instances where servers return well-formed JSON-RPC success responses despite incorrect behavior. No error object is emitted, yet the coordination contract is violated. This mirrors the gray failure phenomenon in distributed systems \cite{huang2017gray}, where systems appear healthy yet deliver degraded service. Similar silent failures have been documented in deep learning systems \cite{humbatova2020taxonomy}.

Although Security faults are less frequent than categories such as Tool Execution, Data~\&~Schema, and State~\&~Configuration, survey respondents consistently rated them as high severity. This divergence suggests that repository frequency and operational risk should be interpreted separately. Security faults may be discovered during private audits, deployment hardening, or internal incident response rather than through public GitHub issue discussions. Prior empirical work on security bug characteristics shows that security bugs often carry high severity. Wei et al.~\cite{wei2021comprehensive} found that, among 1,076 security bugs from five open-source projects, 40.1\% were classified as high severity and 17.9\% as critical severity using CVSS v3.0 base scores. Repository-based frequency counts are insufficient to characterize the significance of security-relevant MCP fault categories.

Finally, initialization-related faults occur across multiple distinct steps in the session startup process (e.g., use of uninitialized state, version mismatches, and missing initialization notifications). No single step accounts for a dominant share of these faults. Because errors occur across several steps rather than at a single point, instability appears inherent in the startup sequence itself, suggesting that correcting a single step is unlikely to eliminate it.

\smallskip
\noindent\textbf{Implications for MCP Server Developers.}
The observed fault patterns motivate five practices tied to specific taxonomy categories and MCP constructs. First, tool inputs should be validated against the declared \texttt{inputSchema} before execution, addressing \emph{Tool Execution} and \emph{Data Type \& Structure} faults associated with \texttt{tools/call}. Second, failures should preserve MCP error semantics: protocol-level failures should use JSON-RPC \texttt{error} responses, whereas tool-execution failures should use MCP tool-result error signaling; this addresses \emph{Message Semantics}, \emph{Tool Execution}, and \emph{Tool Result Propagation}. Third, request handling should be gated on completion of initialization, addressing \emph{Initialization} and \emph{State Transition} faults associated with \texttt{initialize} and \texttt{notifications/initialized}. Fourth, JSON-RPC responses should preserve the \texttt{id} of the corresponding request, addressing \emph{Message Structure} and \emph{Message Semantics} faults involving missing or mismatched identifiers. Fifth, diagnostic output should be kept off \texttt{stdout} in stdio-based servers, addressing \emph{Stdio Transport} faults in which non-protocol output corrupts the MCP message stream~\cite{modelcontextprotocol2025spec}.

The taxonomy supports real-world MCP development in three concrete ways. (1)~\textit{diagnostic triage}, when an MCP server misbehaves, the 27 subcategories provide a structured fault-localization checklist, narrowing investigation to protocol-level hypotheses such as \emph{Schema Compatibility}, \emph{Session State}, and \emph{Tool Result Propagation}. (2)~\textit{Test-suite design}: developers can derive conformance tests directly from leaf categories (e.g., verifying that timeout violations yield structured JSON-RPC error objects and that undeclared tool identifiers are correctly rejected, and that requests are not processed before initialization completes.). (3)~\textit{Risk-based prioritization}: repository frequency and practitioner confirmation should be considered together. Repository frequency identifies categories that appear in public development artifacts, such as Tool faults (192/833, 23.0\%), Data~\&~Schema (147/833, 17.6\%), and State~\&~Configuration (143/833, 17.2\%). Practitioner confirmation further highlights categories that developers recognize in practice, such as Tool Execution (73\%) and Schema Compatibility (69\%). These signals indicate where validation and review efforts should concentrate.

\smallskip
\noindent\textbf{Implications for Protocol and SDK Maintainers.} Several fault classes are better addressed through SDK infrastructure than through repeated fixes in individual implementations.  SDKs can enforce request-response identifier consistency (\emph{Message Structure} and \emph{Message Semantics}), validate declared tool inputs by default (\emph{Tool Execution} and \emph{Data Type \& Structure}), prevent premature request handling (\emph{Initialization} and \emph{State Transition}), and provide helpers for correct protocol- and tool-level error handling (\emph{Message Semantics}, \emph{Tool Execution}, and \emph{Tool Result Propagation}).


Protocol documentation can clarify edge cases around capability declaration, cancellation, progress reporting, and tool-result propagation~\cite{modelcontextprotocol2025spec}. Consolidating these correctness checks into SDKs and protocol guidance could reduce implementation effort and ease compliance across languages and frameworks.

\smallskip
\noindent\textbf{Implications for Benchmarking and Fault Injection.} Each taxonomy subcategory represents an observed fault pattern that can be seeded to evaluate detection, diagnosis, and recovery. Practitioners can inject schema mismatches, missing initialization notifications, malformed tool results, incorrect success responses, or corrupted transport output to assess whether servers, SDKs, and monitors detect the violation. As in deep learning, where mutation operators covered only a fraction of real fault types~\cite{humbatova2020taxonomy}, MCP reliability benchmarks should be grounded in observed protocol-level faults rather than generic software bugs.

\smallskip
\noindent\textbf{Implications for Repository and Issue Management.} Maintainers can use taxonomy categories as issue labels, triage tags, or checklist items during code review, making fault reports easier to search, compare, and resolve. Prior work found that issue management features such as labels and assignees are associated with closure outcomes~\cite{yang2023users}. Our taxonomy provides a starting point for studying whether protocol-aware labeling improves MCP issue triage and resolution.

\smallskip
\noindent\textbf{Implications for Researchers.}  Many confirmed faults yield structurally valid responses while violating coordination semantics, making detection strategies based solely on crashes or explicit errors insufficient. Similar silent failures have been reported in deep learning and hybrid quantum-classical systems~\cite{humbatova2020taxonomy,bensoussan2025taxonomy}. Correctness must therefore be asserted at the behavioral level: oracles should validate response content, session-state consistency, and the occurrence of required notification events across interaction steps. The 27 subcategories define a fault-injection space for evaluating detection and correction capabilities, providing a foundation for research into MCP robustness.

\section{Threats to Validity}
\label{sec:threats}

\smallskip
\noindent\textbf{Internal Validity.}
Internal validity concerns the soundness of the methods and the control of biases in the study design~\cite{wohlin2012experimentation}. Thread reconstruction from GitHub artifacts may be incomplete when issues, pull requests, and commits are inconsistently cross-referenced. To mitigate this, we used automated reference matching to link related artifacts and manually inspected reconstructed threads before labeling. Threads lacking sufficient diagnostic evidence were excluded from the confirmed dataset. Also, Independent coding of 833 threads by two authors may still be influenced by prior MCP knowledge, and threads with incomplete diagnostic context may have been miscoded. All disagreements (52/833, 6.24\%) were resolved through discussion with reference to the MCP specification, and threads lacking sufficient evidence were excluded. Residual misclassification cannot be ruled out, particularly for threads describing failures that span multiple categories.

\smallskip
\noindent\textbf{External Validity.} External validity concerns the generalizability of the findings~\cite{wohlin2012experimentation}. The dataset is restricted to public GitHub repositories passing activity and collaboration filters, so private MCP servers, enterprise deployments, and repositories outside GitHub may exhibit fault patterns absent from this study. We combined two public registries, applied inclusion criteria, and validated the taxonomy with 55 practitioners drawn from MCP repositories to broaden coverage beyond the corpus. The taxonomy characterizes documented MCP server faults and does not cover MCP clients, host applications, or agent orchestration frameworks, whose fault characteristics may differ.

\smallskip
\noindent\textbf{Construct Validity.} Construct validity concerns whether the study accurately operationalizes the concepts it intends to measure~\cite{wohlin2012experimentation}. MCP repositories contain ordinary bugs, build failures, and deployment issues alongside protocol-specific faults, creating a risk that the taxonomy conflates general software defects with MCP runtime failures. We applied a three-stage filtering process, a fault-related keyword filter, an MCP-specific relevance filter (motivated by an 80\% false-positive rate in a 100-artifact pilot inspection), and a final manual review, and iteratively aligned category definitions with the official MCP specification across five coding rounds. Some category boundaries remain interpretive, particularly at the intersection of schema compatibility, tool execution, and tool-result propagation, where the violated constraint can be framed at multiple levels of abstraction.

\smallskip
\noindent\textbf{Conclusion Validity.} Conclusion validity concerns whether the inferences are adequately supported by the collected evidence~\cite{wohlin2012experimentation}. Frequency counts are derived from 833 manually coded threads and should be treated as descriptive evidence of the studied corpus, not as prevalence estimates across the MCP ecosystem. GitHub artifacts reflect only faults that developers choose to report publicly; faults resolved through private channels, internal trackers, or commits without issue links leave no trace, and repository data can be incomplete or biased~\cite{kalliamvakou2016depth,munaiah2017curating}. Additionally, threads that were abandoned or closed without resolution~\cite{li2021you} may have been excluded despite describing real faults, potentially underrepresenting failures that were never publicly resolved. To avoid over-relying on frequency alone, we use the practitioner survey to independently assess perceived severity and diagnostic effort, since a frequently reported category is not necessarily the most operationally severe, and a rarely reported one may still carry high risk.

\section{Conclusion}
\label{sec:conclusion}

MCP is becoming a critical integration layer between LLM-enabled applications and external tools, data sources, and services, yet runtime faults in MCP servers have remained empirically underexplored. We addressed this gap by manually analyzing 833 MCP-specific runtime fault threads from 473 maintained GitHub repositories, deriving a taxonomy of 11 top-level categories, 27 subcategories, and 73 leaf fault types. A survey of 55 MCP server developers validated its practical relevance: all subcategories were observed in practice, and no additional fault class emerged. Faults frequently arise at protocol interaction boundaries, spanning message structure, transport behavior, session lifecycle, schema compatibility, tool execution, and cancellation handling. The taxonomy provides an empirical basis for MCP reliability engineering: developers can use it for fault diagnosis and test design; SDK and protocol maintainers can use it to identify correctness checks for infrastructure; and researchers can use it to construct oracles, benchmarks, and fault-injection studies. As the MCP ecosystem matures, some fault categories may shift in frequency, and new ones may emerge. Future work should examine fault patterns across transport mechanisms, languages, and deployment settings, and extend the analysis to MCP clients, host applications, and agent orchestration frameworks.

\section*{Acknowledgments}

This research was supported by the Safeguard project (grant No.\ 20506020) and received funding from Deloitte. The authors thank the practitioners who participated in the validation survey and provided feedback on the proposed taxonomy.

\section*{Data Availability Statement}
\label{sec:replication}

To support transparency and reproducibility, we provide a replication package containing the mined MCP datasets, manual coding records, and analysis scripts.  It also provides all figures and tables used in the paper, enabling the recreation of the taxonomy, frequency analyses, and validation results.\footnote{\url{https://figshare.com/s/c1e1e802e1b9f38fddf0}}

\section*{Conflict of Interest Statement}

The authors declare no conflicts of interest.

\bibliography{references}

\end{document}